\documentclass[%
 preprint,
amsmath,amssymb,
aps,
pra,  
longbiblioography,
]{revtex4-2}

\usepackage[pdftex, pdftitle={Article}, pdfauthor={Author}]{hyperref}

\usepackage{graphicx}
\usepackage{dcolumn}
\usepackage{bm}

\usepackage{verbatim}
\usepackage{setspace}
\usepackage{multirow}

\usepackage{xcolor}

\begin{document}

\preprint{APS/123-QED}

\title{Global isotopic analysis of hyperfine-resolved rotational spectroscopic data for barium monofluoride, BaF}

\author{Alex Preston, Graceson Aufderheide, Will Ballard, and Richard Mawhorter}
\affiliation{\footnotesize Department of Physics and Astronomy, Pomona College, Claremont, California 91711-6327, USA}

\author{Jens-Uwe Grabow}
\affiliation{\footnotesize Gottfried-Wilhelm-Leibniz-Universität, Institut für Physikalische Chemie and Elektrochemie, Lehrgebiet A, D-30167 Hannover, Germany}

\date{\today}

Prepared for submission to Physical Review A using REVTeX 4.2a

\begin{abstract}
New high-precision microwave spectroscopic measurements and analysis of rotational energy level transitions in the ground vibronic state of the open-shell BaF molecule are reported with the purpose of contributing to studies of physics beyond the Standard Model. BaF is currently among the key candidate molecules being examined in the searches for a measurable electron electric dipole moment, \textit{e}EDM, as well as the nuclear anapole moment. Employing Fourier-transform microwave spectroscopy, these new pure rotational transition frequencies for the $^{138}$Ba$^{19}$F, $^{137}$Ba$^{19}$F, $^{136}$Ba$^{19}$F, $^{135}$Ba$^{19}$F, and $^{134}$Ba$^{19}$F isotopologues are analyzed here in a combined global fit with previous microwave data sets for $^{138}$Ba$^{19}$F ($v=0-4$), $^{137}$Ba$^{19}$F, and $^{136}$Ba$^{19}$F using the program SPFIT. As a result, hyperfine parameters are significantly improved, and we observe a distinctive structure in a Born-Oppenheimer breakdown (BOB) analysis of the primary rotational constant. This can be understood using the nuclear field shifts due to the known isotopic variation in the size of barium nuclei and in combination with the smaller linear mass-dependent BOB terms.
\end{abstract}
\maketitle

\section{Introduction}
\label{sec:Introduction}
Heavy diatomic systems have long been fruitful subjects of studies in atomic, molecular, and optical physics. Among the diatomic molecules currently at the forefront of the field, BaF is a particularly intriguing candidate for studying parity non-conservation (PNC) and the underlying fundamental physics and symmetry involved. In particular, BaF is being examined in studies concerning the determination of the electric dipole moment of the electron ($e$EDM) \cite{Steimle_BaF_Optical, BaF_eEDM_Kozlov(1997), BaF_eEDM_Titov(2005), BaF_eEDM_Nayak(2006), BaF_eEDM_Titov(2006), BaF_eEDM_Nayak(2007), BaF_eEDM_Nayak(2008), BaF_eEDM_Nayak(2009), BaF_eEDM_Kozlov(1995), BaF_eEDM_Aggarwal(2018), BaF_eEDM_Vutha(2018)} and nuclear anapole moments \cite{BaF_Anapole_DeMille(2008), BaF_Anapole_Altunas(2018), BaF_Anapole_Safronova(2018)}.

Toward these ends, in just the last few years several laboratories have made significant progress on laser cooling \cite{Langen_BaF_Serrodyne_Cooling, Comparat_BaF, Yan_BaF_DopplerCooling} and deflecting \cite{Hessels_BaF2023} BaF beams in a variety of ways, including the ability to simultaneously cool and separate beams of different isotopologues \cite{Kogel_BaF_Isotope_Cooling}.  Another significant achievement is the very recent announcement of the first BaF magneto-optical trap (MOT) \cite{Yan_MOT}.  These advances have been based on absorption spectroscopy studies \cite{Langen_etal_BaF, Yan_BaF_SatSpec} in the same time period.

In this work, microwave spectroscopic techniques have been used to measure pure rotational transitions within the ground $X^2\Sigma^{+}$ ($v=0$) state of the five most-abundant naturally occurring isotopologues of BaF (labeled $\alpha=1-5$): $^{138}$Ba$^{19}$F (71.7\%), $^{137}$Ba$^{19}$F (11.2\%), $^{136}$Ba$^{19}$F (7.9\%), $^{135}$Ba$^{19}$F (6.6\%), and $^{134}$Ba$^{19}$F (2.4\%). Data for $^{135}$Ba$^{19}$F and $^{134}$Ba$^{19}$F are reported for the first time in the literature.

Using SPFIT from Pickett's CALPGM suite of programs \cite{SPFIT,Novick_SPFIT_guide,Drouin_SPFIT}, rotational transition frequencies measured in this work were fit together with previously published BaF rotational transition frequencies \cite{Ryz_Tor_BaF, Ryz_etal_BaF, Ernst_etal_BaF}, generating a comprehensive set of effective Hamiltonian parameters used to model the energy level structure. The successful global fit of data for the five present isotopologues offers significant insight into not only the characteristics defining each individual species, but more significantly allows for investigation into the differences in behavior of the even and odd isotopes of barium. Since nuclear anapole moments derive from nuclear spin-dependent parity violation, the sizable non-zero nuclear spin present in both odd isotopes of barium, $\textit{I}\:(^{137,135}{\rm Ba})=3/2$, makes the isotopologues $^{137}$Ba$^{19}$F and $^{135}$Ba$^{19}$F particularly interesting for such anapole analysis. The even isotopologues possess a single nuclear spin, that of the fluorine,  $\textit{I}\:(^{19}{\rm F})=1/2$. Furthermore, the $\textit{I} > 1/2$ nuclear spins of the odd barium nuclei give rise to nuclear electric quadrupole moments \textit{Q} in addition to their intrinsic nuclear magnetic moments. The presence of sizable nuclear electric quadrupole moments, $Q_{137}=+16.0(3)$ fm$^{2}$ and $Q_{135}=+24.5(4)$ fm$^{2}$ \cite{IUPAC}, indicates that the nuclei of the odd barium isotopes are significantly non-spherical. 

In order to harmonize the contrasting even and odd isotopologue properties and obtain a satisfactory global fit of the data from the five varients of BaF studied, Born-Oppenheimer breakdown (BOB) terms were included on the primary rotational constant. This deliberate flexibility beyond rigid mass scaling to the reference isotopologue $^{138}$Ba$^{19}$F ($\alpha=1$) enabled a global fit to this diverse data set within experimental uncertainty. As will be discussed in more detail below, implementing this BOB analysis made for a significant improvement in the quality of the fit alongside the magnetic hyperfine effects of the nuclear spins and electric quadrupole moments \textit{Q} of the odd isotopes of barium. The pronounced structure of the barium odd-even isotope nuclear charge radii over these 5 isotopologues also enabled the untangling of the underlying nuclear size or field shift \cite{Nuclear_Size_Almoukhalalati(2016), Nuclear_Size_Knecht(2012)} and mass-dependent BOB components.  Similar features would also be evident in a molecular King plot \cite{King_Plot_A-K}. Observing and quantifying these detailed nuclear effects lays a foundation for PNC experiments studying EDM's and the nuclear spin-dependent anapole moment. Beyond this, such data help probe other subtle features in BaF \cite{Haase_theory_2020} and even more deformed nuclei in molecules such as RaF\cite{RaF_ExcStateSpec}, e.g. the distribution of magnetization within the nucleus known as the Bohr-Weisskopf effect \cite{NuclMag_RaF}.

\section{Experiment}
\label{sec:Experiment}
BaF rotational spectroscopy data were collected using the Fourier-transform microwave (FTMW) spectrometer housed in the \textit{Institut für Physikalische Chemie und Elektrochemie} at the Gottfried Wilhelm Leibniz Universität in Hannover, Germany \cite{HannoverGroup}. Deviating from
the Balle-Flygare-configuration \cite{Balle-Flygare}, the apparatus operates in the coaxially oriented beam-resonator arrangement (COBRA)  \cite{COBRA_1, COBRA_2, COBRA_3, COBRA_4, COBRA_5}. A molecular jet of BaF is generated by ablating a sample of barium attached to the end of a rotating stainless steel rod with a Nd:YAG pulse laser with 1064 nm fundamental and then frequency doubled to 532 nm. The laser 8 ns pulse is triggered at 20 Hz repetition rate, and each pulse ablates with roughly 20 mJ of energy. A buffer gas of mostly inert neon is introduced to the environment, carrying approximately 2\% CF$_{4}$ by concentration with which the barium reacts to create the desired BaF molecules \cite{Graceson}.

This process occurs in the ablation nozzle on the back-side of one of the reflectors forming the Fabry-Perot-type resonator. When the ablation seeded gas is released into the resonator it undergoes supersonic expansion, passing from $\sim4-5$ bar in the ablation nozzle to $\sim10^{-5}$ mbar in the resonator. This effectively cools the rotational temperature of the molecules to about 2 K, increasing the population in the lowest rotational energy levels while collisional velocity equilibration also results in sub-Doppler line-widths.

\begin{figure*}
    \centering
    \includegraphics[width=\textwidth]{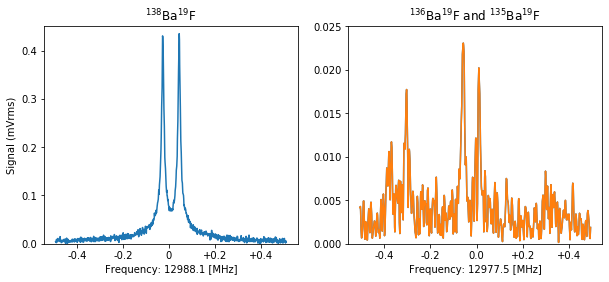}
    \caption{Examples of BaF spectra, including the strong $^{138}$Ba$^{19}$F transition near 12988.1 MHz (the 5th $^{138}$Ba$^{19}$F entry in Table \ref{tab:BaF Data}) and a pair of weaker transitions from $^{136}$Ba$^{19}$F near 12977.2 MHz and $^{135}$Ba$^{19}$F near 12977.5 MHz (the 3rd entry for both isotopologues in Table \ref{tab:BaF Data}).  All three signals exhibit the characteristic double-peak feature as a result of the Doppler shifted components within the resonator; the rest frequency is the arithmetic mean of the two peak frequencies \cite{COBRA_1}.  Experimental conditions were optimized for the central transitions, and the TEM-mode's Gaussian envelope function reduces the relative intensity of the $^{136}$Ba$^{19}$F transition compared to the more central $^{135}$Ba$^{19}$F transition.}
    \label{fig:BaF Data Spectrum}
\end{figure*}

For a thorough description of the COBRA FTMW spectrometer, see \cite{COBRA_1, COBRA_3}. A short ($\sim\ 1$ $\mu$s) microwave pulse is emitted within the tuned resonator forming a standing wave field along the axis of the supersonic expansion, exciting and polarizing the molecular species in the jet.  The system is capable of measuring frequencies in the range of $2-26.5$ GHz.  With strong signal-to-noise frequencies of unblended lines are determined to an accuracy on the order of 0.5 kHz or better, and neighboring transitions separated by more than 6 kHz can be resolved \cite{Mawhorter_PbF, Mawhorter_YbF}. A Fourier transform of the free induction decay (FID) produces a frequency doublet by virtue of the beat frequency from the Doppler effect between the molecular jet and the resonator's standing wave transverse-electric-magnetic (TEM)-mode pattern \cite{COBRA_1}. Example BaF spectra are seen in Figure \ref{fig:BaF Data Spectrum}, featuring the characteristic signal doublets. For this open shell molecule external magnetic fields are nulled out within the cavity by an array of three pairs of Helmholtz coils surrounding the apparatus. These allow enough control of the nulled external and then induced magnetic fields along chosen axes to enable sensitive Zeeman effect studies of other moleccules, such as PbF \cite{PbF_g_factors}.  Here no systematic effects due to stray fields are observed beyond the level of experimental accuracy.

\section{Observations}
\label{sec:Observations}
\subsection{Molecular Structure}\label{subsec:Molecular Structure}

The structure of the $X ^2\Sigma^{+}$ ground state of BaF for low rotational levels is best described by Hund's case (\textit{b}) \cite{Steimle_BaF_Optical} which has helpful visualizations in Figs. 4.6 and 4.7 of Gordy and Cook's comprehensive reference work \cite{GordyandCook}. There exists, however, a slight structural difference between the even (bosonic) and odd (fermionic) isotopologues given the lack of or presence of, respectively, a nuclear spin in the barium isotope.  Ernst, et al. \cite {Ernst_etal_BaF} provide an energy level diagram for the even cases and Kogel, et al. \cite{Langen_BaF_arxiv} provide one for the odd cases.

For the even isotopologues $^{138}$Ba$^{19}$F, $^{136}$Ba$^{19}$F, and $^{134}$Ba$^{19}$F the energy level patterns are well described by the Hund's case ($b_{\beta J}$) limit. Here the coupling between the rotational angular momentum \textbf{N} and the electron spin angular momentum \textbf{S} gives rise to the intermediate quantum number \textit{J}, such that $\textbf{S}+\textbf{N}=\textbf{J}$. Thus, pairs of fine-structure states $J=N\pm1/2$ stem from rotational levels \textit{N} with parity $(-1)^N$. The separation between these split \textit{J} states within a pair is given by $\gamma(N+1/2)$,  \cite{BaF_Albrecht_etal_2020} where $\gamma$ is the electron spin-rotation constant shown in Table \ref{tab:Dunham & Gamma Par} below. Magnetic hyperfine coupling of the fluorine nuclear spin, $\textbf{I}(^{19}{\rm F})=1/2$, splits both fine-structure levels \textit{J} into two further hyperfine levels \textit{F}, each such that the total angular momentum \textit{F} is defined by $\textbf{I}_1(^{19}{\rm F})+\textbf{J}=\textbf{F}$.

As noted above, the structure of the two odd isotopologues $^{137}$Ba$^{19}$F and $^{135}$Ba$^{19}$F with Ba nuclear spins differs from that of the even species. The large $^{137,135}$Ba magnetic hyperfine interaction requires a sequential Hund's case ($b_{\beta S}$) coupling scheme to best describe the ground state of the odd BaF isotopologues for low-rotational levels \cite{Steimle_BaF_Optical, BaF_Albrecht_etal_2020}. In this scheme the good intermediate quantum number \textit{G} arises from the coupling between the electron spin angular momentum \textbf{S} and the nuclear spin angular momentum of the odd barium isotopes $\textbf{I}_1$, such that $\textbf{S}+\textbf{I}_1(^{137,135}{\rm Ba})=\textbf{G}(^{137,135}{\rm Ba})$. Furthermore, there exists an intermediate angular momentum $\textbf{F}_1$, given by $\textbf{N}+\textbf{G}=\textbf{F}_1$. The total angular momentum \textbf{F} in this case emerges from the weak coupling between $\textbf{F}_1$ and the nuclear spin of the fluorine, giving $\textbf{F}_1+\textbf{I}_2(^{19}{\rm F})=\textbf{F}$. The spectra of the odd isotopologues are also more complex than those of their even counterparts due to the hyperfine and rotational splittings being of similar magnitude \cite{BaF_Albrecht_etal_2020}.

\subsection{Data Outline}\label{subsec:Data Outline}

The global fit of BaF microwave spectroscopic data described in this work was achieved by first fitting the previously published data  collated in the NIST diatomic spectral database \cite{NIST} from the following studies: C. Ryzlewicz and T. Törring (1980) \cite{Ryz_Tor_BaF}, C. Ryzlewicz, H.-U. Schütze-Pahlmann, J. Hoeft, and T. Törring (1982) \cite{Ryz_etal_BaF}, and W. E. Ernst, J. Kändler, and T. Törring (1986) \cite{Ernst_etal_BaF}. After successfully combining the data from all of these studies in a single fit, high-precision FTMW transition frequencies measured using the COBRA spectrometer were added in a final global fit to confirm the assignment and to improve the effective Hamiltonian parameters for BaF. Data included in the global fit are summarized in Table \ref{tab:Data in Fit}.

The current data, presented in the Appendix in Table \ref{tab:BaF Data}, contain frequencies corresponding to the $N=1-0$ and $N=2-1$ rotational energy level bands in the ground electronic and vibrational states for the five aforementioned isotopologues. These two rotational bands coincide with frequency values in the vicinity of either 13 GHz or 26 GHz, making these BaF transitions the only two which the COBRA spectrometer can determine given its spectral range of $2-26.5$ GHz. Note that for the even isotopologue data the SPFIT convention $\textbf{F}_1=\textbf{F}$ is used to employ a consistent set of quantum numbers.  The data from \cite{Ryz_Tor_BaF} and \cite{Ryz_etal_BaF} contain higher frequency values, having probed rotational transitions as high as $N=22-21$ with 30 kHz uncertainties. Furthermore, the former study provided the only vibrationally excited transitions in the global fit (up to $v=4$ in the reference isotopologue $^{138}$Ba$^{19}$F). Finally, three of the six frequencies published in \cite{Ernst_etal_BaF} were remeasured using the COBRA spectrometer.

\begin{table}[h]
\resizebox{\textwidth}{!}{
\begin{tabular}{lccccccc} 
\hline\hline
\multicolumn{1}{c}{\textbf{Source}} & \textbf{Isotope} & \begin{tabular}[c]{@{}c@{}}\textbf{Lines}\\\textbf{of}\\\textbf{Data}\end{tabular} & \begin{tabular}[c]{@{}c@{}}\textbf{Freq.}\\\textbf{Range}\\\textbf{(GHz)}\end{tabular} & \begin{tabular}[c]{@{}c@{}}\textbf{Freq.}\\\textbf{Unc.}\\\textbf{($\pm$MHz)}\end{tabular} & \begin{tabular}[c]{@{}c@{}}\textbf{Rotational}\\\textbf{Transitions}\\\textbf{($N=X_n-X_{n-1}$)}\end{tabular} & \begin{tabular}[c]{@{}c@{}}\textbf{Vibration}\\\textbf{Levels}\\\textbf{($v$)}\end{tabular} & \textbf{Hyperfine} \\ 
\hline\hline
\multirow{5}{*}{\begin{tabular}[c]{@{}l@{}}\textbf{This Work}\end{tabular}}                    & $^{138}$Ba$^{19}$F & 10 & 13 and 26 & 0.0005 $\rightarrow$ 0.002 & 1-0 and 2-1 & 0 & Yes \\
& $^{137}$Ba$^{19}$F & 9 & 13 and 26 & 0.0005 $\rightarrow$ 0.001 & 1-0 and 2-1 & 0 & Yes \\
& $^{136}$Ba$^{19}$F & 9 & 13 and 26 & 0.0005 $\rightarrow$ 0.003 & 1-0 and 2-1 & 0 & Yes \\
& $^{135}$Ba$^{19}$F & 9 & 13 and 26 & 0.0005 $\rightarrow$ 0.002 & 1-0 and 2-1 & 0 & Yes \\
& $^{134}$Ba$^{19}$F & 2 & 13 & 0.001 and 0.002 & 1-0 & 0 & Yes \\ 
\hline
\begin{tabular}[c]{@{}l@{}}Ryzlewicz\\ \& Törring \\\end{tabular} & $^{138}$Ba$^{19}$F & 24 & 77 $\rightarrow$ 285 & 0.03 & 6-5 $\rightarrow$ 22-21 & 0 $\rightarrow$ 4 & No \\
\cite{Ryz_Tor_BaF} & $^{136}$Ba$^{19}$F & 5 & 91 $\rightarrow$ 285 & 0.03 & 7-6 $\rightarrow$ 22-21 & 0 & No \\ 
\hline
\begin{tabular}[c]{@{}l@{}}Ryzlewicz\\\textit{et al.}\;\cite{Ryz_etal_BaF}\end{tabular} & $^{137}$Ba$^{19}$F & 24 & 104 and 285 & 0.06 & 8-7 and 22-21 & 0 & Yes \\ 
\hline
\begin{tabular}[c]{@{}l@{}}Ernst\\\textit{et al.}  \cite{Ernst_etal_BaF} \end{tabular} & $^{138}$Ba$^{19}$F & 3 & 13 and 26 & 0.02 & 1-0 and 2-1 & 0 & Yes \\
\hline\hline
\textbf{Total} & \begin{tabular}[c]{@{}c@{}}\textit{Five}\\\textit{Isotopologues}\end{tabular} & 95 & 13 $\rightarrow$ 285 & \textit{Varied} & 1-0 $\rightarrow$ 22-21 & 0 $\rightarrow$ 4 & \textit{Partial} \\
\hline\hline
\end{tabular}}
\caption{BaF spectral data used to perform a global fit. This work introduces the first data for isotopologues $^{135}$Ba$^{19}$F and $^{134}$Ba$^{19}$F.}
\label{tab:Data in Fit}
\end{table}

The inclusion of the first sub-kHz-precision microwave spectroscopy measurements for the low-abundance isotopologues $^{135}$Ba$^{19}$F and $^{134}$Ba$^{19}$F in the FTMW data set offers considerable enhancements to the growing catalog of BaF spectroscopic knowledge. The previous studies dealt with either one or two of the more prevalent isotopologues, as Ref. \cite{Ryz_Tor_BaF} worked with the even isotopologues $^{138}$Ba$^{19}$F and $^{136}$Ba$^{19}$F, Ref. \cite{Ryz_etal_BaF} focused on the odd isotopologue $^{137}$Ba$^{19}$F, and Ref. \cite{Ernst_etal_BaF} concentrated on the reference isotopologue $^{138}$Ba$^{19}$F. Thus, by introducing and analyzing a uniform set of measured frequencies for five isotopologues, this work bolsters the overall inventory of BaF spectroscopic data and enables a detailed investigation of the contrasting physics present in the even and odd isotopologues of BaF.

\section{Analysis}
\label{sec:Analysis}
\subsection{Hamiltonian and Parameters}

A set of rotational and hyperfine parameters were determined from the measured transition frequencies using the SPFIT program, which generated a fit to an effective Hamiltonian for the $X ^2\Sigma^{+}$ state. This effective Hamiltonian specifies the framework from which the spectroscopic model is developed. Due to the non-zero nuclear spin of odd mass isotopes of Ba, additional terms are needed in the effective Hamiltonian for BaF in these cases. For the simpler even Ba-mass BaF isotopologue systems, the effective Hamiltonian for the $X ^2\Sigma^{+}$ state is given by \cite{Ryz_etal_BaF, Steimle_BaF_Optical}:
\begin{equation}
\begin{split}
    \textbf{H}^{eff}(^2\Sigma^+)_{even} = & B\textbf{N}^2-D\textbf{N}^4+\gamma \textbf{N}\cdot\textbf{S} + b_F (\rm{^{19}F})\textbf{I}\cdot\textbf{S} \\& +c(\rm{^{19}F})\left(\textit{I}_z\textit{S}_z-\frac{1}{3}\textbf{I}\cdot\textbf{S}\right) + \textit{C}_I(\rm{^{19}F})\textbf{I}\cdot\textbf{N}
\label{eq:Even Hamiltonian}
\end{split}
\end{equation}
where \textbf{N} is the angular momentum of the nuclear frame excluding the electronic spin \textbf{S} and nuclear spin \textbf{I}. Here $I_z$ and $S_z$ are projections along the molecular axis and only \textbf{I}($^{19}$F) is considered in this case since the even barium isotopes have no nuclear spin. 

In this work the primary rotational constant $B$ and centrifugal distortion constant $D$, as well as the spin-rotation coupling constant $\gamma$, are depicted in expanded Dunham notation since the vibrationally excited transitions of the reference isotopologue $^{138}$Ba$^{19}$F allow for the resolution of vibrationally sensitive spectroscopic parameters. The conventional spectral coefficients are equivalent to the following Dunham expansions:
\begin{equation}
    B_v=Y_{01}+Y_{11}\left(v+\frac{1}{2}\right)+Y_{21}\left(v+\frac{1}{2}\right)^2+...
\end{equation}
\begin{equation}
    D_v=Y_{02}+Y_{12}\left(v+\frac{1}{2}\right)+...
\end{equation}
The spin-rotation coupling constant $\gamma$ is expressed by three parameters in this work: $\gamma_e$, its vibrational dependence $\alpha_{\gamma}$, and the centrifugal distortion $\delta_{\gamma}$, here denoted $\gamma_{00}$, $\gamma_{10}$, and $\gamma_{01}$.  Furthermore, of the two magnetic hyperfine structure Frosch-Foley parameters $b$ and $c$, the former is portrayed instead as the physically-significant Fermi-contact parameter $b_F=b+c/3$. The nuclear spin-rotation coupling constant $C_I$ is also included in this work.

Since the odd isotopes of barium do possess nuclear spin, their effective Hamiltonian is more complicated, requiring additional barium hyperfine parameters \cite{Ryz_etal_BaF, Steimle_BaF_Optical}:
\begin{equation}
\begin{split}
    \textbf{H}^{eff}(^2\Sigma^+)_{odd} = & \textbf{H}^{eff}(^2\Sigma^+)_{even} + b_F(^{137,135}\rm{Ba})\textbf{I}\cdot\textbf{S} \\& + c(^{137,135}\rm{Ba})\left(\textit{I}_z\textit{S}_z - \frac{1}{3}\textbf{I}\cdot\textbf{S}\right) \\& + eQq_0(^{137,135}\rm{Ba}) \frac{3\textit{I}_z - \textbf{I}^2}{4\textit{I}(2\textit{I}-1)} \\& + \textit{C}_I(^{137,135}\rm{Ba})\textbf{I}\cdot\textbf{N}
\label{eq:Odd Hamiltonian}
\end{split}
\end{equation}
The fluorine hyperfine parameters in Equation \ref{eq:Even Hamiltonian} apply to all 5 BaF isotopologues since they all share the single fluorine isotope ($^{19}$F). 

\subsection{Fitting the Data}

In order to ensure consistent treatment of the data from the 4 different studies in Table \ref{tab:Data in Fit}, five individual fits were first made to all of the data for each isotopologue, often spanning 2 or more studies. Consistent with covariant methods, SPFIT assigns each data point a weight according to the inverse square of the stated uncertainty. As described in detail in Ref. \cite{Novick_SPFIT_guide}, for the 1$\sigma$ statistics employed here an RMS fit quality parameter of 1 indicates that approximately 68\% of the observed - calculated values are appropriately within the stated uncertainty for the set, be it larger or smaller.   

Of the five single-isotopologue fits, the parameters obtained for the reference isotopologue $^{138}$Ba$^{19}$F were the best determined due to the dominant natural abundance of $^{138}$Ba$^{19}$F which provides a comparative factor of 6 or more in signal. $^{138}$Ba$^{19}$F data also contained the only vibrationally excited transitions (v = 0-4), allowing the determination of the vibrationally sensitive coefficients. Furthermore, $^{138}$Ba$^{19}$F (index $\alpha=1$) data enabled the best determination of the fluorine hyperfine parameters $b_F$(F), $c$(F), and $C_I$(F). The values for the vibration-dependent coefficients in the remaining isotopologue fits (indices $\alpha=2,3,4,5$) were isotopically scaled based on reduced mass ratios \cite{Drouin_Scaling}, while the fluorine hyperfine parameters were held constant to those obtained by the $^{138}$Ba$^{19}$F fit. The barium hyperfine parameters $b_F$(Ba), $c$(Ba), $eQq_0$(Ba), and $C_I$(Ba) were then introduced in the two odd isotopologue fits.

Following the results of the five individual fits, separate even ($\alpha=1,3,5$) and odd ($\alpha=2,4$) fits were produced, with the odd case proving more challenging. The odd-even staggering of the nuclear charge radii required granting further BOB flexibility to the leading rotational constant. 
Additionally, the $eQq_0$(Ba) parameter is scaled by the nuclear electric quadrupole moment $Q$, and the $\sim$2\% uncertainty in the standard ratio of these values \cite{IUPAC} is so much larger than the FTMW frequency uncertainties that, initially, the $^{135}$Ba$^{19}$F and $^{137}$Ba$^{19}$F $eQq_0$ values had to be floated individually. Nevertheless the $eQq_0$ value for each of the two odd isotopologues optimized such that their ratio agreed within error with the original scaled ratio, resulting in both consistent and much-improved $eQq_0$(Ba) values.

Combining these even and odd isotopologue fits into a global fit of this rich set of microwave spectroscopic data for all five stable isotopologues of BaF ultimately resulted in a satisfactory RMS value of 1.083. 

\subsection{Born-Oppenheimer Breakdown Implementation}

A Born-Oppenheimer breakdown (BOB) analysis is often required when working with multiple isotopologues of diatomics (see Table 6 in Ref. \cite{Etchison_BOB_SrS}, and \cite{ABC_Article, PbF_Article}). Several effects can contribute to BOB residuals, and for a heavy polar diatomic like BaF at this level of precision the two most important of these are the field shift effect due to the differences in nuclear size among the barium isotopes and a mass-dependent effect accounting for the tiny, but no longer negligible, ratio between the electron and nuclear masses. These were first described together and compared in the groundbreaking 1982 work of Schlemback and Tiemann \cite{Schlembach_Tiemann}, and an analysis following their approach is employed here.

For BaF it was adequate to provide additional BOB flexibility for the $Y_{01}$ parameters of the four alpha isotopologues, as originally suggested by LeRoy \cite{LeRoy_BOB}. 

Since fluorine has only one stable isotope, for BaF the BOB-corrected $Y_{01}$ value for each isotopologue $X'$ relative to the reference isotopologue $X=^{138}$BaF is given in the Schlembach-modified Watson model \cite{Watson1980, Serafin2007} as
\begin{equation}\label{eq:Y01massdep}
    Y_{01}\,=\,\mu^{-1}\overline{U}_{01}\left[ 1 + m_e\frac{\Delta ^X_{01}}{M_{X'}} + V_X \delta\langle r^2 \rangle_{X X'}\right]
\end{equation}
where $\mu$ is the reduced mass of the appropriate BaF isotopologue, $m_e$ is the electron mass, $M_{X'}$ is the atomic mass of the substituted isotope, $\Delta_{01}$ is a unitless Watson-type mass-dependent BOB correction, and $V_X\delta\langle r^2 \rangle_{XX'}$ is the field shift term that depends on the change in mean square charge radii between the two nuclei $X$ and $X'$.  The $\overline{U}_{01}$ term is:

\begin{equation}
    \overline{U}_{01}\,=\, U_{01}\left( 1 + V_X\langle r^2 \rangle_X\right)
\end{equation}
where $U_{01}$ is the isotope-independent rotational spectral parameter, $U_{01} = \mu Y_{01}$, and the $\langle r^2 \rangle$ term is the mean square charge radius of the reference nucleus $X$. Further discussion of mean square charge radii and one table of their $\delta\langle r^2 \rangle$ difference values are given in \cite{Angeli2013}.\\

The very small ($\lesssim$ 1 kHz) BOB offset correction terms were implemented as 1 or 2 sets of additive terms for either or both effects to the isotopically scaled $Y_{01}$ parameter for each of the 4 nonreference isotopologues $Y_{01}^{\alpha}$. In 3 separate cases a fit was used to determine $V_X$ (setting $\Delta_{01}^{X}=0$), $\Delta_{01}^{X}$ (setting $V_X=0$), and the best combination of these using Equation \ref{eq:Y01massdep} above following the general procedure outlined in Bizzocchi, et al. \cite{Bizzocchi_etal}. Here first the reference isotopologue dependent BOB $\delta_{01}^X$ parameter is determined and the corresponding equivalent unitless, reference independent $\Delta_{01}^X$ value is then calculated. This analysis is presented in more detail in Jackson \textit{et al.} \cite{PbF_Article}.

\begin{table*}[t]
\resizebox{\textwidth}{!}{
\centering
\begin{tabular}{ldddddddd}
\hline\hline
\\[-1em]
Parameters [MHz] 
& \multicolumn{1}{c}{$Y_{01}$}
& \multicolumn{1}{c}{$Y_{11}$}
& \multicolumn{1}{c}{$10^2Y_{21}$} 
& \multicolumn{1}{c}{$10^3Y_{02}$}
& \multicolumn{1}{c}{$10^6Y_{12}$}
& \multicolumn{1}{c}{$\gamma_e$}
& \multicolumn{1}{c}{$10^2\alpha_{\gamma}$} 
& \multicolumn{1}{c}{$10^4\delta_{\gamma}$} \\
\hline
\\
\textit{Fit Parameters}
\\ \\
$^{138}$Ba$^{19}$F
& \textbf{6491}.\textbf{396\;062(168)}
& \textbf{-34}.\textbf{883\;266(300)}
& \textbf{1}.\textbf{596(28)}
& \textbf{-5}.\textbf{525\;23(53)}
& \textbf{-9}.\textbf{12(78)}
& \textbf{80}.\textbf{915\;550(250)}
& \textbf{-5}.\textbf{31(72)}
& \textbf{1}.\textbf{666(134)}
\\ \\
\textit{Derived Parameters}
& 
& 
& 
& 
& 
& 
& 
& 
\\ \\
$^{137}$Ba$^{19}$F
& 6497.133\;423(307)
& -34.929\;527(300)
& 1.599(28)
& -5.535\;00(53)
& -9.14(78)
& 80.987\;072(250)
& -5.32(72)
& 1.669(134)
\\
$^{136}$Ba$^{19}$F
& 6502.966\;469(284)
& -34.976\;576(301)
& 1.602(28)
& -5.544\;94(53)
& -9.16(78)
& 81.059\;781(250)
& -5.33(72)
& 1.672(134)
\\
$^{135}$Ba$^{19}$F
& 6508.871\;622(427)
& -35.024\;231(301)
& 1.605(28)
& -5.555\;02(53)
& -9.18(79)
& 81.133\;393(251)
& -5.33(72)
& 1.675(135)
\\
$^{134}$Ba$^{19}$F
& 6514.879\;191(409)
& -35.072\;731(302)
& 1.608(28)
& -5.565\;28(53)
& -9.20(79)
& 81.208\;276(251)
& -5.34(72)
& 1.678(135)
\\ \\
\hline\hline
\\[-1em]
\multicolumn{1}{c}{Parameters [MHz]}
& \multicolumn{1}{c}{$Y_{01}$}
& \multicolumn{1}{c}{$Y_{11}$}
& \multicolumn{1}{c}{$10^2Y_{21}$}
& \multicolumn{1}{c}{$10^3Y_{02}$}
& \multicolumn{1}{c}{$10^6Y_{12}$}
& \multicolumn{1}{c}{$\gamma_e$}
& \multicolumn{1}{c}{$10^2\alpha_{\gamma}$}
& \multicolumn{1}{c}{$10^4\delta_{\gamma}$}
\\
\hline
\\
\textit{Literature Parameters}
\\ \\
$^{138}$Ba$^{19}$F
& 6491.396\;2(11)^a
& -34.883\;1(11)^a
& 1.593 (22)^a
& -5.525\;0(11)^a
& -9.43(95)^a
& 80.984(19)^a
& -5.84(73)^a
& 1.12(17)^a
\\ 
$^{137}$Ba$^{19}$F
& & & & & & & 
& 1.13(147)^b
\\
$^{136}$Ba$^{19}$F
& 6502.966\;05(63)^a
& -34.976\;4(11)^a
& 1.599(22)^a
& -5.544\;8(11)^a
& -9.47(95)^a
& 81.152(43)^a
& -5.84(73)^a
& 1.12(17)^a
\\
$^{135}$Ba$^{19}$F
& & & & & & & & 
\\
$^{134}$Ba$^{19}$F
& & & & & & & & 
\\
\hline\hline
\\[-1em]
\multicolumn{1}{c}{Parameters [MHz]}
& \multicolumn{1}{c}{$B_0$}
& \multicolumn{1}{c}{}
& \multicolumn{1}{c}{}
& \multicolumn{1}{c}{$-10^3D_0$}
& \multicolumn{1}{c}{}
& \multicolumn{1}{c}{$\gamma_0$}
\\
\hline
\\
\textit{Literature Parameters}
\\ \\
$^{138}$Ba$^{19}$F
& 6474.045(66)^e
&
& 
& -5.539\;0(54)^e
& 
\\
$^{137}$Ba$^{19}$F
& 6479.677\;3(32)^b
&
& 
& -5.544(3)^b
& 
& 80.923(6)^c 
\\
$^{136}$Ba$^{19}$F 
& 
& 
& 
\\
$^{135}$Ba$^{19}$F
& 6498.0(15)^d 
& 
& 
\\
$^{134}$Ba$^{19}$F 
&
&
&
\\
\hline\hline
\end{tabular}}
\caption{\textit{Top}: Spectral Dunham coefficients and $\gamma$ spin-rotation constants obtained by the fit with 1$\sigma$ uncertainties. Parameters that are freely floated are in \textbf{bold}. Note the uncertainties on the derived $Y_{01}$ parameters differ notably from that of the reference isotopologue by virtue of the inclusion of a field shift and a Born-Oppenheimer breakdown mass shift on this parameter. \textit{Middle}: Previously published Dunham coefficients and $\gamma$ spin-rotation constants for BaF. \textit{Bottom}: The alternative classical spectral notation parameters as published in \cite{Ryz_etal_BaF, Ernst_etal_BaF, Steimle_BaF_Optical, Barrow_BaF}. \\
\footnotesize
$^a$ \cite{Ryz_Tor_BaF} \;
$^b$ \cite{Ryz_etal_BaF} \;
$^c$ \cite{Ernst_etal_BaF} \;
$^d$ \cite{Steimle_BaF_Optical} (wavenumber converted to MHz) \;
$^e$\cite{Barrow_BaF} (wavenumber converted to MHz)}
\label{tab:Dunham & Gamma Par}
\end{table*}

\begin{table*}[t]
\begin{ruledtabular}
\begin{tabular}{lcddd}
\\[-1em]
\multicolumn{1}{c}{Parameters [MHz]} & \multicolumn{1}{c}{Isotope} &
\multicolumn{1}{c}{\textbf{This Work}} & \multicolumn{1}{c}{Previous Experiment} &
\multicolumn{1}{c}{Theory} \\
\hline
\\[-1em]
\multicolumn{1}{c}{$b$~(Ba)~($=A_{\perp}$)} & 137 & 2302.56(102)^a & 2301.(9)^c & 2305.(130)^b \\
& 135 & 2060.07(92)^a & & \\
\multicolumn{1}{c}{$c$~(Ba)} & 137 & \textbf{75}.\textbf{197\;9(57)} & 75.(6)^c & 78.(4)^b \\
& 135 & 67.278\;7(53) & & \\
\multicolumn{1}{c}{$b_F$~(Ba)~($=b+c/3$)} & 137 & \textbf{2327}.\textbf{63(101)} & 2326.(11)^c & 2321.(140)^b \\
& 135 & 2082.50(91) & 2104.(15)^e & \\
\multicolumn{1}{c}{$b+c$~($=A_{\parallel}$)} & 137 & 2377.76(102)^a & 2376.(15)^c & 2383.(130)^b \\
& 135 & 2127.35(92)^a & & \\
\multicolumn{1}{c}{$eQq_0$~(Ba)} & 137 & \textbf{-143}.\textbf{680(12)} & -117.(12)^c & \\
& 135 & \textbf{-93}.\textbf{405(12)} & & \\
\multicolumn{1}{c}{$10^3C_I$~(Ba)} & 137 & \textbf{1}.\textbf{70(42)} & & \\
& 135 & 1.52(38) & & \\
\hline
\multicolumn{1}{c}{$b$~(F)} & 19 & 63.414\;3(9)^a & 63.509(32)^d & \\
& & & 60.(6)^c & \\
\multicolumn{1}{c}{$c$~(F)} & 19 & \textbf{7}.\textbf{305\;4(13)} & 8.224(58)^d & \\
\multicolumn{1}{c}{$b_F$~(F)~($=b+c/3$)} & 19 & \textbf{65}.\textbf{849\;4(5)} & 66.250(32)^d & \\
\multicolumn{1}{c}{$10^3C_I$~(F)} & 19 & \textbf{26}.\textbf{13(17)} & & \\
\end{tabular}
\end{ruledtabular}
\caption{Hyperfine parameters of fluorine and barium obtained by the fit of this work as compared to previous experimental and theoretical results. Parameters that are freely floated are in \textbf{bold}. Due to the single isotope of fluorine the same values of $b_F$(F), $c$(F), and $C_I$(F) are used for all isotopologues of BaF. The hyperfine parameters of barium, however, apply only to the odd isotopes of the element. Note that the nuclear spin-rotation parameters $C_I$ for the non-zero spin fluorine and barium nuclei are determined here for the first time. \\ 
\footnotesize
$^a$ Calculated from $b_F$ and $c$ \\
$^b$ \cite{Haase_theory_2020} ($c$ and $b_F$ calculated; $c$ error from 5.5\% $A_{\perp}$ and $A_{\parallel}$ error) \\ $^c$ \cite{Ryz_etal_BaF} ($b_F$ calculated) \\ $^d$ \cite{Ernst_etal_BaF} ($b_F$ calculated) \\ $^e$ \cite{Steimle_BaF_Optical} (cm$^{-1}$ results converted to MHz)}
\label{tab:Hyperfine Params}
\end{table*}

\begin{table*}
\scriptsize
\begin{ruledtabular}
\begin{tabular}{cddddd}
Fits
& \multicolumn{1}{c}{\begin{tabular}[c]{@{}c@{}}FINAL\\ Fit\end{tabular}}
& \multicolumn{1}{c}{\begin{tabular}[c]{@{}c@{}}Only\\ Field Shift\end{tabular}}
& \multicolumn{1}{c}{\begin{tabular}[c]{@{}c@{}}Only\\ Mass Shift\end{tabular}} 
& \multicolumn{1}{c}{\begin{tabular}[c]{@{}c@{}}Offset\\ All Float\end{tabular}} 
& \multicolumn{1}{c}{\begin{tabular}[c]{@{}c@{}}No\\ Shifts\end{tabular}} \\ 
\hline
Shift Terms & \multicolumn{1}{c}{2} & \multicolumn{1}{c}{1} & \multicolumn{1}{c}{1} & \multicolumn{1}{c}{4} & \multicolumn{1}{c}{0} \\
\hline
\\[-1em]
RMS Error & \multicolumn{1}{c}{1.083} & \multicolumn{1}{c}{1.086} &   \multicolumn{1}{c}{1.099} & \multicolumn{1}{c}{1.051} & \multicolumn{1}{c}{1.174} \\
\hline\hline \\[-1em]
\textit{$Y_{01}$ offsets} [kHz] & & & & &
\\ \\[-1em]
$^{137}$Ba$^{19}$F
& -0.48(37)
& -0.57(14)
& -0.17(4)
& -0.62(20)
& 
\\
$^{136}$Ba$^{19}$F
& -0.41(22)
& -0.39(9)
& -0.33(8)
& -0.27(12)
& 
\\
$^{135}$Ba$^{19}$F
& -0.73(39)
& -0.76(18)
& -0.50(13)
& -0.92(26)
& 
\\
$^{134}$Ba$^{19}$F
& -0.61(25)
& -0.51(12)
& -0.66(17)
& -1.46(46)
& 
\\ \\[-1em]
\hline\hline \\[-1em]
\textit{Mass Shift (BOB)} & & & & &\\ 
 \\[-1em]
$\delta_{01}^{Ba}$ [kHz]
& 7.8(102)
& 
& 22.5(57)
& 48.(16)
& 
\\ \\[-1em]
$\Delta_{01}^{Ba}$
& -0.30(39)
& 
& -0.87(22)
& -1.9(6)
& 
\\ \\[-1em]
\hline\hline \\[-1em]
\textit{Field Shift} [fm$^{-2}$] & & & & &\\ 
 \\[-1em]
$V_{Ba}$
& 10.6(60)\times10^{-7}
& 14.3(34)\times10^{-7}
&
& 
&
\\
\end{tabular}
\end{ruledtabular}
\caption{Comparison of the $Y_{01}$ offset values and derived mass shift BOB and field shift terms for the non-reference alpha isotopologues ($\alpha=2-5$) under various constraints, ranging from only strict isotopic scaling to all 4 offsets freely floating. To give the clearest sense of the information inherent in the data, fractional uncertainties of the corresponding offset parameters determined in the fit are used here for the field shift terms $V_{Ba}$ and mass shift BOB delta terms $\Delta_{01}^{Ba}$ and $\delta_{01}^{Ba}$ (not to be confused with the $\delta\langle r^2 \rangle$ term from the nuclear physics literature).  As visualized in Figure \ref{fig:Mass and Field Shift Plot}, the field shift term plays a larger role in improving the fit.} 
\label{tab:Mass and Field Shift Table}
\end{table*}

\begin{figure*}
    \centering
    \includegraphics[width=\textwidth]{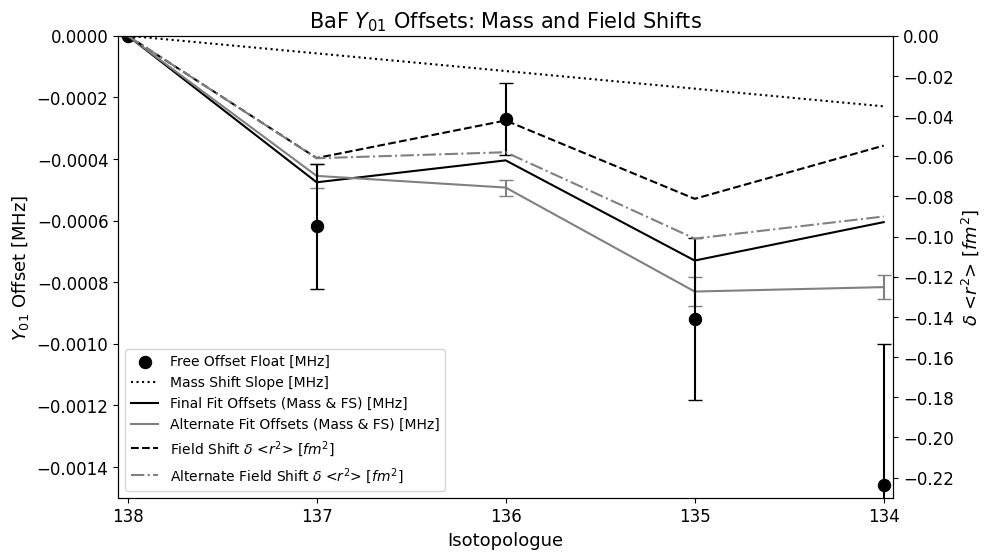}
    \caption{Visualization of the $Y_{01}$ offsets in the final fit (black solid line) as the sum of mass shift BOB (dotted) and field shift (dashed) contributions. The black data points represent the free floating offsets and associated 1$\sigma$ uncertainties from Table \ref{tab:Mass and Field Shift Table}. Note that the staggered trend of the field shift values (scaled on the right axis using the linear relationship in Eq. \ref{eq:Y01massdep}}) dominates the effect on the $Y_{01}$ offsets in the final fit. The more negative Landolt-Börnstein $\delta\langle r^2 \rangle$ field shifts \cite{Landolt-Bornstein} are included (in gray dot-dash) on the same scale for comparison with the recently posted BaF King plot data \cite{Langen_BaF_arxiv}.  To facilitate approximate comparison with our data the same mass shift has been added to these and the $\delta \langle r^2 \rangle$ error bars have been doubled, and improved 2$\sigma$ agreement is evident.
    \label{fig:Mass and Field Shift Plot}
\end{figure*}

\section{Discussion}
\label{sec:Discussion}

Although molecules like BaF are among the most sensitive physical systems for measuring the extremely small ($\sim$ Hz and mHz) effects of parity non-conservation, identifying such effects with confidence requires a thorough understanding of the many different and subtle physical interactions present in such open shell molecules. The best way to provide such a solid foundation is to combine the best and broadest spectroscopic results available in a global fit requiring well-known scaling consistency across the existing isotopologues.  The global fit presented here includes sub-kHz precision FTMW data for all five BaF isotopologues as well as data from a wide variety of excited rotational and vibrational states. The FTMW data provides the ability to distinguish new from established features, and the global nature of this large and diverse data set enables small but important adjustments to parameters describing these interactions from earlier fits of less complete data sets. 

The comparison of the Dunham expansion coefficients $Y_{lm}$ and the expanded electron spin-rotation $\gamma$ parameters in Table \ref{tab:Dunham & Gamma Par} illustrates a case in point. Agreement between the results of this work and those of Refs. \cite{Steimle_BaF_Optical, Ryz_Tor_BaF, Ryz_etal_BaF, Ernst_etal_BaF} for the Dunham expansion coefficients is excellent and for the spin-rotation values are comparable. Considering the different number of Dunham coefficients included, we also find broad agreement with those determined by Guo \textit{et al}. \cite{Bernath}

To clarify the notation in the previous literature (see Ref. \cite{Ryz_etal_BaF}), $\gamma_0=\gamma_{00}+\frac{1}{2}\gamma_{10}=\gamma_e+\frac{1}{2}\alpha_{\gamma}$, and for the centrifugal term $\gamma_{01}=\delta_{\gamma}$, as already noted. Comparing center $\gamma_0$ values from the 3 available studies for the reference and most abundant isotopologue $^{138}$Ba$^{19}$F, we find 80.955(20) MHz from Ref. \cite{Ryz_Tor_BaF}, 80.923(6) MHz from Ref. \cite{Ernst_etal_BaF}, and 80.890(1) MHz for the current global fit. These vary more than their quoted 1$\sigma$ uncertainties here, but are consistent with each other in light of the different parameter sets utilized in their determination.

The hyperfine parameters are presented in Table \ref{tab:Hyperfine Params}. The fluorine hyperfine splittings were not resolved in Ref. \cite{Ryz_Tor_BaF}, were only incompletely determined in Ref. \cite{Ryz_etal_BaF}, and while a focus of Ref. \cite{Ernst_etal_BaF} they were still determined from narrowly separated thus partially blended lines. So the sizeable changes in the Frosch-Foley terms $b_F$(F) and $c$(F) in the present work are somewhat anticipated, given the newly added $C_I$(F) and $C_I$(Ba) nuclear spin-rotation terms and the improved Ba hyperfine terms for the odd isotopologues.

Turning to them, one can see excellent agreement for the $^{137}$Ba$^{19}$F Frosch-Foley terms in the fit from Ref. \cite{Ryz_etal_BaF} and the current global fit.  The values for $^{135}$Ba$^{19}$F in the global fit are constrained to the ratio of magnetic moments for these odd nuclei. The $b_F$($^{135}$Ba) value from the optical study in Ref. \cite{Steimle_BaF_Optical} comes from a fit with many fixed constraints and is in very good agreement, only 1.4$\sigma$ above the global fit value.

For this $^2\Sigma$ molecule the relationships between $b$ and $c$ and the theoretically calculated electron spin resonance (ESR) $\bf{A}$ tensor parameters $A_{||}$ and $A_{\perp}$ are quite simple \cite{Mawhorter_PbF} . Once unpacked it is clear that agreement is also quite good for the recent theoretical calculations of Haase, et al. \cite{Haase_theory_2020}, and indeed is much better than the stated conservative error estimates. To provide a direct comparison with our experimentally determined parameters $b_F$(Ba) and $c$(Ba) in Table \ref{tab:Hyperfine Params} related parameters are calculated as indicated and the original literature parameters are also provided.  

The $C_I$(Ba) terms in the global fit are also constrained to use the ratio of magnetic moments as well as the appropriate reduced mass scaling. The values are quite a bit smaller than the single $C_I$(F) value, but are well-determined.

The largest change is in the $^{137}$Ba$^{19}$F $eQq_0$ nuclear electric quadrupole coupling constant, which is just over 2$\sigma$ larger than the value from Ref. \cite{Ryz_etal_BaF}, which itself has a 10\% uncertainty. Here the $^{135}$Ba$^{19}$F $eQq_0$ value was determined independently, and since these values scale with the nuclear electric quadrupole moment \textit{Q}, their ratio can be used as a consistency check. This ratio from the global fit is 0.6501(1), which is right between and in excellent agreement with both the IUPAC \textit{Q} ratio \cite{IUPAC} of 0.653(12) and the more recent value in Pyykkö's end of 2017 compilation of 0.648(12) \cite{Pyykko_Qvalues_2018}. Besides the smaller changes previously noted, the provision of a $^{135}$Ba$^{19}$F $eQq_0$ value from high precision $^{135}$Ba$^{19}$F microwave data as well as a significantly different and much improved $^{137}$Ba$^{19}$F $eQq_0$ value are among the main spectroscopic contributions of this work.

Data for multiple BaF isotopologues helps to determine the BOB effects in this open shell molecule. From Table \ref{tab:Mass and Field Shift Table} we can see that as expected, the best fit at RMS 1.051 is afforded by allowing each isotopologue $Y_{01}$ value to vary freely. At the other end of the scale, rigidly enforcing isotopic scaling (no BOB correction) results in a much worse RMS of 1.174, due in part to systematic residuals in the highest precision FTMW data. The usual approach of refining a single mass-dependent $\Delta_{01}$ BOB correction value applied to all isotopologues brings the RMS down to 1.099, whereas the RMS using a $V_X$ field shift only approach is lower at 1.086. This is often the case for diatomics containing heavier elements where the field shift effect is larger because the physically relevant underlying structure of the isotopic nuclear charge radii plays a role. A final fit adding the $\Delta_{01}$ term back in brings the RMS down a bit further to 1.083. The RMS values shown in Table \ref{tab:Mass and Field Shift Table} provide a sense of how well the whole data set responds to the different BOB models employed here.

The interesting physics appears in Figure \ref{fig:Mass and Field Shift Plot}, which shows as a function of barium isotope the direct offsets amounting to only about 1 kHz between the freely floating refined $Y_{01}$ values for each case and the corresponding isotopically scaled $Y_{01}$ parameters. One expects a generally downward trend starting from $^{138}$BaF for two reasons. First, the $\Delta_{01}$ term is negative, and second, nuclei will be getting smaller moving down from the most abundant heaviest reference isotope , i.e. $\delta\langle r^2 \rangle$ is generally negative. Even more noteworthy is how well the odd-even staggering of the nuclear charge radii $\delta\langle r^2 \rangle$ values \cite{Angeli2013} track the observed offsets, as their values are plotted to scale on the right hand side of the graph. 

A second set of $\delta\langle r^2 \rangle$ values incorporating higher moments \cite{Landolt-Bornstein} from a Landolt-Börnstein (L-B) reference work that are used in the very recently posted molecular King plot \cite{King_Plot_A-K} for BaF \cite{Langen_etal_BaF} are also shown on the same scale in Figure \ref{fig:Mass and Field Shift Plot}. The Angeli set is a recalculated version of data from the L-B set \cite{Angeli2013}.  Nevertheless, the excellent  agreement with the independent spectroscopic data shown in Ref. \cite{Langen_etal_BaF} and the much improved agreement for our $^{135}$BaF and especially $^{134}$BaF offsets (now comfortably within 2$\sigma$) indicate a strong preference for the L-B set.  To provide an approximate overall comparison, the sum of these with the plotted mass shift values is shown with doubled $\delta\langle r^2 \rangle$ error bars. 

Since the mass-dependent term is always essentially linear and most nuclear size differences for other heavy atoms are more uniformly monotonic, this distinctive odd-even nuclear size structure is very helpful in beginning to tease apart the usually quite correlated nuclear size and mass dependent effects. Here for BaF the graph visualizes that the mass dependent part is about 1/3 of the combined effect. Although with large uncertainties for these sub-kHz effects, numerically the field shift term $V_{Ba}$ drops $\sim$ 1/3 from near 15 fm$^{-2}$ when $\delta_{01}^{Ba}=0$ to about 10 fm$^{-2}$ in the combined fit and $\delta_{01}^{Ba}$ drops from approximately $\sim$ 23 kHz when $V_{Ba}=0$ to the vicinity of $\sim$ 8 kHz in the combined fit. 

Although the Greek root of \textit{barium} denotes being \textit{heavy}, $^{208}$Pb is much heavier than $^{138}$Ba, and the BOB analysis in our recent study of PbF isotopologues \cite{PbF_Article} provides an interesting comparison case. For PbF the observed offsets are about 8 times larger than for BaF, and in terms of the field shift parameters themselves,V$_{Pb}$ is about 4.5  times larger than V$_{Ba}$ reported here. BOB analyses have been carried out for a number of PbX diatomics, and a comparison shows that for PbF the nuclear field shift effect dominates at about 90\% of the combined BOB effect. As the mass-dependent effect is smaller for heavier atoms, using the field shift effect alone worked well for PbF, where this single BOB parameter fit matched the small deviation from linearity for $^{207}$PbF and even smaller deviation for $^{206}$PbF within much smaller relative experimental uncertainty.

It isn't currently possible to reliably calculate $\Delta_{01}$ values for open shell molecules \cite{Knecht_Saue}, which otherwise would offer another way to break up the correlation. So in this context the combination of the opportunity to begin separating the effects and the ensuing reasonable result for middle weight barium is quite instructive. 

Very few systems of comparable molecules which have a similar isotope structure of stable nuclei have been studied in such detail, with the possibly notable exception of hafnium.  Hf also has two relatively abundant odd nuclei, and although their \textit{Q}-values are a factor of 15-20 larger than for the odd isotopes of barium the hafnium nuclear charge radii vary fairly smoothly. We have carried out a preliminary BOB study on existing data of closed shell HfO \cite{Lesarri_HfO, Lovas_HfO} and find a similar smooth behavior in the $Y_{01}$ offsets as expected from the field shift model. Nevertheless the HfO offsets themselves are much larger than those observed for heavier PbF, so some puzzles remain.

Ytterbium has two odd isotopes and only one with a nuclear quadrupole moment (\textit{Q}($^{173}$Yb) = 280 fm$^2$), however this is quite large and a study of YbF could provide another interesting case between BaF and PbF. We are in the process of combining FTMW data on the even isotopologues of YbF with earlier studies of $^{171}$YbF and $^{173}$YbF\cite{171YbF, 173YbF}.

\section{Conclusion}
\label{sec:Conclusions}

New sub-kHz precision FTMW data of the five stable isotopologues of the diatomic radical BaF presented here have been analyzed in conjunction with data from three earlier microwave studies which also include transitions of highly-excited rotational and vibrational states. The resulting global fit significantly improves our knowledge of the mass-scaled rotational constants $Y_{vJ}$ and hyperfine parameters, especially the spin-rotation parameter $\gamma$, the fluorine Frosch-Foley terms $b_F$(F) and $c$(F), and the nuclear electric quadrupole coupling constant $eQq_0$ for the odd isotopologues. New information is provided for the isotopologues $^{135}$BaF and $^{134}$BaF and the nuclear spin-rotation terms $C_I$ for both barium and fluorine are refined for the first time.

Successfully fitting the global data set requires a Born-Oppenheimer breakdown analysis of the leading Dunham expansion coefficient $Y_{01}$. This is shown to arise from a combination of the nuclear size field shift effect and a smaller mass-dependent effect consistent with the barium isotope's nuclear charge radii and masses. The large number of barium isotopes and the odd-even staggering of these radii greatly facilitates a better understanding of these often heavily-correlated BOB components. Thus, the BaF BOB analysis here serves as a link between the much heavier Pb$Y$ ($Y=$ Se, Te, S, O, F) and Tl$Y$ ($Y=$ F, Cl, Br, I) molecules \cite{Giuliano_2008,Schlembach_Tiemann, Serafin2007, PbF_Article} and lighter molecules such as AlCl \cite{ABC_Article} where the mass-dependent effect (proportional to 1/M) dominates. The marked consistency of our BOB data with the recently posted BaF King plot \cite{Langen_etal_BaF} also sheds light on differing Ba nuclear radii $\delta\langle r^2 \rangle$ values.

Beyond this, interest in BaF is increasing as a probe of fundamental physics. Examples include the design of parity non-conservation experiments whose ultimate goals include the experimental measurement of the electron electric dipole moment (eEDM) as well as the nuclear anapole moment. 
Relating spectral observations, once made, to actual values of these quantities relies on the quantum chemical prediction of nonobservables, e.g. for the eEDM effective electric fields experienced by an unpaired electron. The validity of such predictions can, in part, be probed by their power to also reproduce other molecular field-dependent quantities, such as the very precise hyperfine parameters obsutained here. Other interesting future work could include a theoretical study of the odd-even BOB terms as well as the further investigation of molecules whose heavy atoms have a similar isotope structure, such as Ti\cite{Witsch_TiO}, Mo, Ru, Nd, Sm, Gd, and Dy\cite{Lasner_DyO}.

\section*{Acknowledgements}
\label{sec:acknowledgements}

The authors thank Sven Herbers and Sean Jackson for technical assistance, and Tim Langen, Tim Steimle, and Trevor Sears for helpful discussions.  For the Pomona College authors funding has been provided by a Pomona College Sontag Fellowship as well as a Hirsch Research Initiation grant and the Summer Undergraduate Research Program. and for J-UG support from the DFG GR1344/11-1 is gratefully acknowledged.




\section*{Appendix}
\label{sec:appendix}

A table of frequencies and SPFIT quantum numbers is shown in Table \ref{tab:BaF Data}.

\begin{table}
\scriptsize
\centering
\begin{ruledtabular}
\begin{tabular}{l|ccccccccc|dd} 
\multicolumn{1}{c|}{}   & $N'$ & $J'$ & $F_1'$ & $F'$ &  & $N''$ & $J''$ & $F_1''$ & $F''$ & \multicolumn{1}{c}{Obs. (MHz)} & \multicolumn{1}{c}{Obs.-Calc. (MHz)}  \\ 
\hline\hline
\multirow{10}{*}{$^{138}$Ba$^{19}$F} & 1   & 1/2 & 1     & 1/2  &  & 0    & 1/2  & 1      & 1/2   & 12836.6440 & 0.0005 \\
& 1   & 1/2 & 0     & 0    &  & 0    & 1/2  & 1 & 1/2   & 12864.5445 & 0.0003 \\
& 1   & 1/2 & 1     & 1/2  &  & 0    & 1/2  & 0 & 0     & 12902.4928 & -0.0001 \\
& 1   & 3/2 & 1     & 1/2  &  & 0    & 1/2  & 1      & 1/2   & 12954.0562 & -0.0002 \\
& 1   & 3/2 & 2     & 3/2  &  & 0    & 1/2  & 1      & 1/2   & 12988.1089 & 0.0000 \\
& 1   & 3/2 & 1     & 1/2  &  & 0    & 1/2  & 0      & 0     & 13019.9063 & 0.0005 \\
& 2   & 3/2 & 2     & 3/2  &  & 1    & 3/2  & 2      & 3/2   & 25702.9717 & 0.0016 \\
& 2   & 5/2 & 2     & 3/2  &  & 1    & 3/2  & 2      & 3/2   & 25902.8232 & -0.0016 \\
& 2   & 5/2 & 3     & 5/2  &  & 1    & 3/2  & 2      & 3/2   & 25936.0108 & 0.0002 \\
& 2   & 5/2 & 2     & 3/2  &  & 1    & 3/2  & 1      & 1/2   & 25936.8771 & -0.0002 \\
\hline
\multirow{9}{*}{$^{137}$Ba$^{19}$F} & 1 & 1/2 & 2 & 5/2 & & 0 & 1/2 & 1 & 3/2 & 12943.6645 & 0.0008 \\
& 1 & 1/2 & 2 & 3/2 & & 0 & 1/2 & 1 & 1/2 & 12945.9174 & 0.0011 \\
& 1 & 3/2 & 1 & 3/2 & & 0 & 1/2 & 1 & 3/2 & 12950.5072 & -0.0002 \\
& 1 & 3/2 & 3 & 5/2 & & 0 & 1/2 & 2 & 3/2 & 13006.5716 & -0.0012 \\
& 2 & 3/2 & 3 & 7/2 & & 1 & 1/2 & 2 & 5/2 & 25899.4392 & -0.0008 \\
& 2 & 3/2 & 3 & 5/2 & & 1 & 3/2 & 1 & 3/2 & 25906.4189 & 0.0010 \\
& 2 & 5/2 & 2 & 5/2 & & 1 & 1/2 & 2 & 3/2 & 25910.6856 & -0.0002 \\
& 2 & 5/2 & 4 & 9/2 & & 1 & 3/2 & 3 & 7/2 & 25958.7703 & -0.0002 \\
& 2 & 5/2 & 4 & 7/2 & & 1 & 3/2 & 3 & 5/2 & 25959.3899 & -0.0023 \\
\hline
\multirow{9}{*}{$^{136}$Ba$^{19}$F} & 1 & 1/2 & 1 & 1/2 & & 0 & 1/2 & 1 & 1/2 & 12859.5642 & 0.0003 \\
& 1 & 1/2 & 1 & 1/2 & & 0 & 1/2 & 0 & 0 & 12925.4130 & -0.0003 \\
& 1 & 3/2 & 1 & 1/2 & & 0 & 1/2 & 1 & 1/2 & 12977.1595 & 0.0005 \\
& 1 & 3/2 & 2 & 3/2 & & 0 & 1/2 & 1 & 1/2 & 13011.2287 & 0.0001 \\
& 1 & 3/2 & 1 & 1/2 & & 0 & 1/2 & 0 & 0 & 13043.0095 & 0.0011 \\
& 2 & 3/2 & 2 & 3/2 & & 1 & 1/2 & 1 & 1/2 & 25900.4523 & 0.0011 \\
& 2 & 3/2 & 1 & 1/2 & & 1 & 1/2 & 0 & 0 & 25902.4375 & -0.0007 \\
& 2 & 5/2 & 3 & 5/2 & & 1 & 3/2 & 2 & 3/2 & 25982.1779 & 0.0004 \\
& 2 & 5/2 & 2 & 3/2 & & 1 & 3/2 & 1 & 1/2 & 25983.0512 & 0.0001 \\
\hline
\multirow{9}{*}{$^{135}$Ba$^{19}$F} & 1 & 1/2 & 2 & 5/2 & & 0 & 1/2 & 1 & 3/2 & 12965.4292 & -0.0014 \\
& 1 & 3/2 & 1 & 3/2 & & 0 & 1/2 & 1 & 1/2 & 12971.3725 & 0.0006 \\
& 1 & 1/2 & 2 & 3/2 & & 0 & 1/2 & 1 & 3/2 & 12977.4755 & -0.0008 \\
& 1 & 3/2 & 3 & 5/2 & & 0 & 1/2 & 2 & 5/2 & 12988.6094 & 0.0022 \\
& 1 & 3/2 & 3 & 7/2 & & 0 & 1/2 & 2 & 5/2 & 13020.9845 & 0.0009 \\
& 1 & 3/2 & 3 & 5/2 & & 0 & 1/2 & 2 & 3/2 & 13029.5162 & 0.0002 \\
& 2 & 3/2 & 3 & 7/2 & & 1 & 1/2 & 2 & 5/2 & 25945.3247 & -0.0017 \\
& 2 & 5/2 & 2 & 5/2 & & 1 & 3/2 & 1 & 3/2 & 25960.4876 & 0.0019 \\
& 2 & 5/2 & 4 & 7/2 & & 1 & 3/2 & 3 & 5/2 & 26004.7374 & -0.0045 \\
\hline
\multirow{2}{*}{$^{134}$Ba$^{19}$F} & 1 & 1/2 & 1 & 1/2 & & 0 & 1/2 & 0 & 0 & 12949.0084 & -0.0029 \\
& 1 & 3/2 & 2 & 3/2 & & 0 & 1/2 & 1 & 1/2 & 13035.0306 & -0.0012 \\
\end{tabular}
\end{ruledtabular}
\caption{BaF FTMW frequencies measured using the COBRA spectrometer broken down by isotopologue. Associated quantum numbers and obs.-calc.values specified by the SPFIT program. Note all transitions exist in the ground vibrational state $v=0$.}
\label{tab:BaF Data}
\end{table}

\begin{thebibliography}{10}

\bibitem{Steimle_BaF_Optical}
T. C. Steimle, S. Frey, A. Le, D. DeMille, D. A. Rahmlow, and C. Linton, \textit{Phys. Rev. A} \textbf{84}, 012508 (2011).

\bibitem{BaF_eEDM_Kozlov(1997)}
M. G. Kozlov, A. V. Titov, N. S. Mosyagin, and P. V. Souchko, \textit{Phys. Rev. A} \textbf{56}, R3326 (1997).

\bibitem{BaF_eEDM_Titov(2005)}
A. V. Titov, N. S. Mosyagin, A. N. Petrov, and T. A. Isaev, \textit{Int. J. Quantum Chem.} \textbf{104}, 223 (2005).

\bibitem{BaF_eEDM_Nayak(2006)}
M. K. Nayak and R. K. Chaudhuri, \textit{J. Phys. B} \textbf{39}, 1231 (2006).

\bibitem{BaF_eEDM_Titov(2006)}
A. V. Titov, N. S. Mosyagin, A. N. Petrov, T. A. Isaev, and D. P. DeMille, \textit{Prog. Theor. Chem. Phys.} \textbf{15}, 253 (2006).

\bibitem{BaF_eEDM_Nayak(2007)}
M. K. Nayak, R. K. Chaudhuri, and B. P. Das, \textit{Phys. Rev. A} \textbf{75}, 022510 (2007).

\bibitem{BaF_eEDM_Nayak(2008)}
M. K. Nayak and R. K. Chaudhuri, \textit{Phys. Rev. A} \textbf{78}, 012506 (2008).

\bibitem{BaF_eEDM_Nayak(2009)}
M. K. Nayak and B. P. Das, \textit{Phys. Rev. A} \textbf{79}, 060502 (2009).

\bibitem{BaF_eEDM_Kozlov(1995)}
M. G. Kozlov and L. N. Labzowsky, \textit{J. Physics B} \textbf{28}, 1933 (1995).

\bibitem{BaF_eEDM_Aggarwal(2018)}
P. Aggarwal, H. L. Bethlem, A. Borschevsky, M. Denis, K. Esajas, P. A. B. Haase, Y. Hao, S. Hoekstra, K. Jungmann, T. B. Meijkneckt, M. C. Mooij, R. G. E. mermans, W. Ubachs, L. Willmann, and A. Zapara, \textit{Eur. Phys. J. D} \textbf{72}, 197 (2018).

\bibitem{BaF_eEDM_Vutha(2018)}
A. C. Vutha, M. Horbatsch, and E. A. Hessels, \textit{Phys. Rev. A} \textbf{98}, 032513 (2018).

\bibitem{BaF_Anapole_DeMille(2008)}
D. DeMille, S. B. Cahn, D. Murphree, D. A. Rahmlow, and M. G. Kozlov, \textit{Phys. Rev. Lett.} \textbf{100}, 023003 (2008).

\bibitem{BaF_Anapole_Altunas(2018)}
E. Altunas, J. Ammon, S. B. Cahn, and D. DeMille, \textit{Phys. Rev. Lett.} \textbf{120}, 142501 (2018).

\bibitem{BaF_Anapole_Safronova(2018)}
M. S. Safronova, D. Budker, D. DeMille, D. F. J. Kimball, A. Derevianko, and C. W. Clark, \textit{Rev. Mod. Phys.} \textbf{90}, 025008 (2018).

\bibitem{SPFIT}
H. M. Pickett, \textit{J. Mol. Spectrosc.} \textbf{148}, 371-377 (1991).

\bibitem{Novick_SPFIT_guide}
S. E. Novick, \textit{J. Mol. Spectrosc.} \textbf{329}, 1-7 (2016).

\bibitem{Drouin_SPFIT}
B. J. Drouin, \textit{J. Mol. Spectrosc.} \textbf{340}, 1-15 (2017).

\bibitem{Drouin_Scaling}
B. J. Drouin, C. E. Miller, H. S. P. Müller, and E. A. Cohen, \textit{J. Mol. Spectrosc.} \textbf{205}, 128-138 (2001).

\bibitem{IUPAC}
I. Mills and the International Union of Pure and Applied Chemistry. Physical and Biophysical Chemistry Division, \textit{Quantities, Units, and Symbols in Physical Chemistry}, 3rd ed. (RSC Pub., Cambridge, UK, 2007).

\bibitem{Pyykko_Qvalues_2018}
P. Pyykkö, \textit{Mol. Phys.} \textbf{116}, 1328-1338 (2018).

\bibitem{Giuliano_2008}
B.M. Giuliano, l. Bizzocchi, S, Cooke, D. Banser, M.Hess, J. Fritzsche, and J.-U. Grabow, \textit{Phys. Chem. Chem. Phys.} \textbf{10}, 2078 (2008).

\bibitem{Nuclear_Size_Almoukhalalati(2016)}
A. Almoukhalalati, A. Shee, and T. Saue, \textit{Phys. Chem. Chem. Phys.} \textbf{18}, 15406 (2016).

\bibitem{Nuclear_Size_Knecht(2012)}
S. Knecht and T. Saue, \textit{Chem. Phys.} \textbf{401}, 103-112 (2012).

\bibitem{King_Plot_A-K}
M. Athanasakis-Kaklamanakis, S. G. Wilkins, A. A. Breier and G. Neyens, \textit{Phys. Rev. X} \textbf{13}, 011015 (2023).

\bibitem{Langen_etal_BaF}
M. Rockenhäuser, F. Kogel, E. Pultinevicius and T. Langen \textit{Phys. Rev. A} \textbf{108}, 062812 (2023).

\bibitem{Langen_BaF_arxiv} F. Kogel, Y. Chamorro, M. Bhattarai, M. Rockenhäuser, T. Garg, D. DeMille, A. Borschevsky and T. Langen, "High-resolution spectroscopy of barium monofluoride:
Odd isotopologues, hyperfine structure and isotope shifts" (2025) arXiv:2506.10940 [physics.atom-ph].

\bibitem{NuclMag_RaF}
L. V. Skripnikov, \textit{J. Chem. Phys.} \textbf{153}, 114114 (2020).

\bibitem{Balle-Flygare}
T. J. Balle and W. H. Flygare, \textit{Rev. Sci. Instrum.} \textbf{52}, 33 (1981).

\bibitem{HannoverGroup}
Institut für Physikalische Chemie und Elektrochemie, Gottfried Wilhelm Leibniz Universität, Hannover, Germany. \url{https://www.pci.uni-hannover.de/}

\bibitem{COBRA_1}
J.-U. Grabow, W. Stahl, and  H. Dreizler, \textit{Rev. Sci. Instrum.} \textbf{67}, no. 12, 4072–4084 (1996).

\bibitem{COBRA_2}
U. Andresen, H. Dreizler, J.-U. Grabow, and W. Stahl, \textit{Rev. Sci. Instrum.} \textbf{61}, 3694  (1990).

\bibitem{COBRA_3}
J.-U. Grabow, E. S. Palmer, M. C. McCarthy, and P. Thaddeus, \textit{Rev. Sci. Instrum.} \textbf{76}, 093106 (2005).

\bibitem{COBRA_4}
J.-U. Grabow and W. Caminati, “Chapter 14 - Microwave Spectroscopy: Experimental Techniques”, in \textit{Frontiers of Molecular Spectroscopy}, edited by J. Laane (Elsevier, Amsterdam, 2009), pp. 383–454.

\bibitem{COBRA_5}
W. Caminati and J.-U. Grabow, “Chapter 15 - Microwave Spectroscopy: Molecular Systems”, in \textit{Frontiers of Molecular Spectroscopy}, edited by J. Laane (Elsevier, Amsterdam, 2009), pp. 455–552.

\bibitem{Graceson}
G. Aufderheide, B.S. thesis, Pomona College, 2020.

\bibitem{Mawhorter_PbF}
R. J. Mawhorter, B. S. Murphy, A. L. Baum, T. J. Sears, T. Yang, P. M Rupasinghe, C. P. McRaven, N. E. Shafer-Ray, L. D. Alphei, and J.-U. Grabow, \textit{Phys. Rev. A} \textbf{84}, 022508 (2011). 

\bibitem{Mawhorter_YbF}
Z. Glassman, R. J. Mawhorter, J.-U. Grabow, A. Le, and T. C. Steimle, \textit{J. Mol. Spectrosc.} \textbf{300}, 7-11 (2014).

\bibitem{BaF_Albrecht_etal_2020}
R. Albrecht, M. Schwarwaechter, T. Sixt, L. Hofer, and T. Langen, \textit{Phys. Rev. A} \textbf{101}, 013413 (2020).

\bibitem{NIST}
NIST Diatomic Spectral Database, \url{https://physics.nist.gov/cgi-bin/MolSpec/dipeiodic.pl}

\bibitem{Ryz_Tor_BaF}
C. Ryzlewicz and T. Törring, \textit{Chem. Phys.} \textbf{51}, 329-334 (1980).

\bibitem{Ryz_etal_BaF}
C. Ryzlewicz, H.-U. Schütze-Pahlmann, J. Hoeft, and T. Törring, \textit{Chem. Phys.} \textbf{71}, 389-399 (1982).

\bibitem{Ernst_etal_BaF}
W. E. Ernst, J. Kändler, and T. Törring, \textit{J. Chem. Phys.} \textbf{84}, 4769 (1986).

\bibitem{Barrow_BaF}
C. Effantin, A. Bernard, J. d'Incan, G. Wannous, J. Vergès, and R. F. Barrow, \textit{Mol. Phys.} \textbf{70}, 735-745 (1990).

\bibitem{Bernath}
B. Guo, Q. Zhang, and P. F. Bernath, \textit{J. Mol. Spectrosc.} \textbf{170}, 59-74 (1995).

\bibitem{LeRoy_BOB}
R. J. LeRoy, \textit{J. Mol. Spectrosc.} \textbf{194}, 189-196 (1999).

\bibitem{Etchison_BOB_SrS}
K. C. Etchison, C. T. Dewberry and S. A. Cooke, \textit{Chem. Phys.} \textbf{342}, 71-77 (2007).

\bibitem{ABC_Article}
A. Preston, S. Jackson and R. Mawhorter, \textit{Chem. Phys. Lett.} \textbf{807} 140089 (2022)

\bibitem{PbF_Article}
S. Jackson, L. Kim, A. Biekert, A. Nguyen, R. J. Mawhorter, T. J. Sears, L. V. Skripnikov, V. V. Baturo, A. N. Petrov, and J.-U. Grabow, \textit{Phys. Rev. A} \textbf{110}, 042808 (2024).

\bibitem{Schlembach_Tiemann}
J. Schlembach and E. Tiemann, \textit{Chem. Phys.} \textbf{68}, 21-28 (1982).

\bibitem{Watson1980}
J. K. Watson, \textit{J. Mol. Spectrosc.} \textbf{80}, 411 (1980).

\bibitem{Serafin2007}
M. Serafin, S. Peebles, C. Dewberry, K. Etchison, G. Grubbs, R. Powoski and S. Cooke, \textit{Chem. Phys. Lett.} \textbf{449}, 33 (2007).

\bibitem{Angeli2013}
I. Angeli and K. Marinova, \textit{Atom. Data Nucl. Data} \textbf{99}, 69 (2013).

\bibitem{Landolt-Bornstein}
G. Fricke, K. Heilig, Landolt-Börnstein: Num. Data and Funct. Relat. in Science and Tech., in: New Series, Group I: Elem. Part., Nuclei and Atoms, vol. 20, Springer Verlag, 2004, 324 pages.

\bibitem{PbF_g_factors}
V. V. Baturo, P. M. Rupasinghe, T. J. Sears, R. J. Mawhorter, J.-U. Grabow and A. N. Petrov, \textit{Phys. Rev. A} \textbf{104}, 012811 (2021).

\bibitem{Bizzocchi_etal}
L. Bizzocchi, B. M. Giuliano, M. Hess and J.-U. Grabow, \textit{J. Chem. Phys.} \textbf{126}, 114305 (2007).

\bibitem{Knecht_Saue}
S. Knecht and T. Saue, \textit{Phys. Chem. Chem. Phys.} \textbf{18}, 15406 (2016).

\bibitem{Yan_BaF_SatSpec}
W. Bu, Y. Zhang, Q. Liang, T. Chen and B. Yan, \textit{Front. Phys.} \textbf{17} 62502 (2022).

\bibitem{Yan_BaF_DopplerCooling}
Y. Zhang, Z. Zeng, Q. Liang, W. Bu and B. Yan, \textit{Phys. Rev. A} \textbf{105}, 033307 (2022).

\bibitem{Hessels_BaF2023}
A. Marsman, M Horbatsch and E. A. Hessels, \textit{Phys. Rev. A} \textbf{108}, 012811 (2023).

\bibitem{Comparat_BaF}
T. Courageux, A. Cournol, D. Comparat, B. Viaris de Lesegno and H. Lignier, \textit{New J. Phys.} \textbf{24}, 025007 (2022). 

\bibitem{Yan_MOT}
Z. Zeng, S. Deng, S. Yang and B. Yan, \textit{Phys. Rev. Lett.} \textbf{133}, 143404 (2024).

\bibitem{Langen_BaF_Serrodyne_Cooling}
M. Rockenhäuser, F. Kogel, T. Garg, S.A. Morales-Ramirez and T. Langen, \textit{Phys. Rev. Research} \textbf{6}, 043161 (2024).

\bibitem{Kogel_BaF_Isotope_Cooling}
F. Kogel, T. Garg, M. Rockenhäuser, S.A. Morales-Ramirez and T. Langen, \textit{New J. Phys.} \textbf{27}, 013001 (2025).

\bibitem{RaF_ExcStateSpec}
M. Athanasakis-Kaklamanakis, et al., \textit{Phys. Rev. A} \textbf{110}, L010802 (2024).

\bibitem{173YbF}
H. Wang, A.T. Le, T.C. Steimle, E.C. Koskelo, G. Aufderheide, R. Mawhorter and J.-U. Grabow, \textit{Phys. Rev. A} \textbf{100}, 022516 (2019).

\bibitem{171YbF}
Z. Glassman, R. Mawhorter, J.-U. Grabow, A. Le and T.C. Steimle, \textit{J. Mol. Spectrosc.} \textbf{300}, 7-11 (2014).

\bibitem{Haase_theory_2020}
P. A. B. Haase, E. Eliav, M. Ilias, and A. Borschevsky, \textit{J. Phys. Chem. A} \textbf{124}, 3157-3169 (2020).

\bibitem{Lesarri_HfO}
A. Lesarri, R. D. Suenram and D. Brugh, \textit{J. Chem. Phys.} \textbf{117}, 9651-9662 (2002).

\bibitem{Lovas_HfO}
R. D. Suenram, F. J. Lovas, G. T. Fraser and K. Matsumura, \textit{J. Chem. Phys.} \textbf{92}, 4724-4733 (1990).

\bibitem{GordyandCook}
W. Gordy and R. L. Cook, Microwave Molecular Spectra, (John Wiley and Sons, New York 1984) 3rd edition, ISBN 0-471-08681-9

\bibitem{Witsch_TiO}
D. Witsch, A. A. Breier, E. Döring, K. M. T. Yamada, T. F. Giesen and G. W. Fuchs, \textit{J. Mol. Spectrosc.} \textbf{377}, 111439 (2021).

\bibitem{Lasner_DyO}
Z. Lasner, N. Albright, K. Rice, J. Doyle and B. Augenbraun, "Spectroscopy of dysprosium monoxide for optical cycling, cooling, and trapping" Poster presented at DAMOP, 2025 Portland, Oregon.


\end{thebibliography}
\end{document}